\documentclass[letter]{aa} 

\usepackage{graphicx}
\usepackage{txfonts}
\newcommand{\HI}{\mbox{H{\sc i}\,}}

\begin{document} 
    
\title{The atomic gas of star-forming galaxies at z$\sim$0.05 as revealed by the Five-hundred-meter Aperture Spherical Radio Telescope}
\subtitle{}
\author{Cheng Cheng\inst{1,2,3},
        Edo Ibar\inst{4},
        Wei Du\inst{2,3},
        Juan Molina\inst{5,6,7},
        Gustavo Orellana-Gonz\'ales\inst{8, 9},
        Bo Zhang\inst{2},
        Ming Zhu\inst{2},
        Cong Kevin Xu\inst{1,2,3},
        Shumei Wu\inst{1,2,3},
        Tianwen Cao\inst{1,2,3},
        Jia-Sheng Huang\inst{1,2,3},
        Roger Leiton\inst{9},
        Thomas M. Hughes\inst{1,4,10,11},
        Chuan He\inst{1,2,3},
        Zijian Li\inst{1,2,3},
        Hai Xu\inst{1,2,3},
        Y. Sophia Dai\inst{1,2,3},
        Xu Shao\inst{1,2,3},
        Marat Musin\inst{1,2,3}
        }
\institute{
        Chinese Academy of Sciences South America Center for Astronomy, National Astronomical Observatories, CAS, Beijing 100101, China.\\
        \email{chengcheng@nao.cas.cn}
        \and
        National Astronomical Observatories, Chinese Academy of Sciences (NAOC), 20A Datun Road, Chaoyang District, Beijing 100101, China
        \and
        CAS Key Laboratory of Optical Astronomy, National Astronomical Observatories, Chinese Academy of Sciences, Beijing 100101, China 
        \and
        Instituto de F\'isica y Astronom\'ia, Universidad de Valpara\'iso, Avda. Gran Breta\~na 1111, Valpara\'iso, Chile.
        \and
        Kavli Institute for Astronomy and Astrophysics, Peking University, 5 Yiheyuan Road, Haidian District, Beijing 100871, P.R. China 
        \and
        Department of Astronomy, School of Physics, Peking University, Beijing 100871, China
        \and
        Departamento de Astronom\'ia (DAS), Universidad de Chile, Casilla 36-D, Santiago, Chile
        \and
        Departamento de Matem\'atica y F\'isica Aplicadas, Universidad Cat\'olica de la Sant\'isima Concepci\'on, Alonso de Ribera 2850, Concepci\'on, Chile
        \and
        Departamento de Astronom\'ia, Universidad de Concepci\'on, Casilla 160-C, Concepci\'on, Chile
        \and
        CAS Key Laboratory for Research in Galaxies and Cosmology,
Department of Astronomy, University of Science and Technology of
China, Hefei 230026, China
    \and
     School of Astronomy and Space Science, University of Science and
Technology of China, Hefei 230026, China
        }

\titlerunning{\HI detection of low-$z$ star-forming galaxies by FAST}
\authorrunning{Cheng et al.}

\abstract
{
We report new \HI observations of four $z\sim 0.05$ VALES galaxies
undertaken during the commissioning phase of the Five-hundred-meter Aperture Spherical Radio Telescope (FAST).}
{
FAST is the largest single-dish telescope in the world, with a 500 meter aperture and a 19-Beam receiver. Exploiting the unprecedented sensitivity provided by FAST, we aim to study the atomic gas content, via the \HI\,21\,cm emission line, in low-$z$ star formation galaxies taken from the Valpara\'iso ALMA/APEX Line Emission Survey (VALES). Together with previous Atacama Large Millimeter/submillimeter Array (ALMA) CO($J=1-0$) observations, the \HI data provides crucial information to measure the gas mass and dynamics.}
{
As a pilot \HI  galaxy survey, we targeted four local star-forming galaxies at $z\sim0.05$. In particular, one of them has already been detected in \HI  by the Arecibo Legacy Fast ALFA survey (ALFALFA), allowing a careful comparison. We use an ON-OFF observing approach that allowed us to reach an rms of 0.7\,mJy\,beam$^{-1}$ at a 1.7\,km\,s$^{-1}$ velocity resolution within only 20 minutes ON-target integration time.
}
{In this letter, we demonstrate the great capabilities of the FAST 19-beam receiver for pushing the detectability of the \HI emission line of extra-galactic sources. The \HI  emission line detected by FAST shows good consistency with the previous Arecibo telescope ALFALFA results. Our observations are put in context with previous multi-wavelength data to reveal the physical properties of these low-$z$ galaxies. We find that the CO($J=1-0$) and \HI  emission line profiles are similar. The dynamical mass estimated from the \HI data is an order of magnitude higher than the baryon mass and the dynamical mass derived from the CO observations, implying that the mass probed by dynamics of \HI is dominated by the dark matter halo. In one case, a target shows an excess of CO($J=1-0$) in the line centre, which can be explained by an enhanced CO($J=1-0$) emission induced by a nuclear starburst showing high velocity dispersion.
}
{}
{}


\maketitle
%

\section{Introduction}

Atomic neutral hydrogen gas, \HI, is found to be one of the most extended baryon components of galaxies \citep{1994Natur.372..530Y}. The width of the \HI  emission line has been historically used to estimate the dynamical mass via the large-scale rotation velocity derived from the double horn shape of the emission line \citep{1978AJ.....83.1026R}. These estimates provide a proxy for estimating the dark matter content in galaxies \citep{2019A&ARv..27....2S}. On the other hand, the optically thin nature of the \HI  emission leads to the possibility of probing possible gas inflows via the asymmetry of the line profile \citep{Bournaud2005, 2020MNRAS.495.1984D}. The \HI fluxes are found to follow very well scaling relations such as: the star formation rate (SFR) surface density $\Sigma_{\rm SFR}$ and the combined surface density of molecular (H$_2$) and \HI gas, $\Sigma_{\rm H{\sc I}  + H_2}$ \citep{1959ApJ...129..243S, Kennicutt1998}, \HI  mass  vs. stellar mass \citep{Huang2012, Maddox2015}, \HI  mass vs.\ \HI  size \citep{2016MNRAS.460.2143W, Stevens2019} and the \HI-to-H$_2$ ratio as a function of stellar or gas surface density \citep{Leroy2008}. These relations enable us to the revelation of important information even when we are unable to perform spatially resolved \HI  observations \citep{Giovanelli2015}.

Given the fundamental importance of the \HI  component, several wide-field \HI  surveys have been carried out to probe the neutral atomic gas at low redshift. Previous blind \HI  surveys such as the \HI  Parkes All-Sky Survey \citep[HIPASS,][13 beams]{2001MNRAS.322..486B, 2004MNRAS.350.1195M, 2006MNRAS.371.1855W}, the \HI  Jodrell All Sky Survey \citep[][4 beams]{2003MNRAS.342..738L}, the Effelsberg-Bonn \HI  Survey \citep[EBHIS,][7 beams]{2010ApJS..188..488W, 2011AN....332..637K}, and the Arecibo Legacy Fast ALFA Survey \citep[ALFALFA, ][7 beams]{2005AJ....130.2598G} have detected large numbers of gas rich galaxies in both northern and southern skies. The on-going Widefield ASKAP L-band Legacy All-sky Blind surveY \citep[WALLABY, ][]{2020arXiv200207311K}, and Deep Investigation of Neutral Gas Origins \citep[DINGO, ][]{2009pra..confE..15M} are planning to perform an \HI  survey over much larger areas using the Australian Square Kilometer Array Pathfinder (ASKAP), expecting to detect  a million new \HI sources up to $z\sim0.4$. 

Current facilities have been able to detect the \HI  emission line up to $z=0.1$ \citep[e.g., ][]{Giovanelli2015}. Major factors limiting the detectability of higher redshift sources include the sensitivity,
the detector frequency range and Radio Frequency Interference (RFI). One way to overcome the sensitivity issues is by using gravitational magnification. For example, 
HI ultra-deep survey projects such as Blind Ultra Deep \HI  Environmental Survey \citep[BUDHIES][]{BUDHIES}, or the COSMOS \HI  Large Extragalactic Survey (CHILES) have spend hundreds hours to extend the \HI  detection up to $z\simeq 0.3$ \citep[$z$ = 0.37 in][ and $z$ = 0.32 in \citealt{2018MNRAS.473.1879R}]{2016ApJ...824L...1F}. Higher redshift \HI  studies have been available from damped Ly$\alpha$ absorbers \citep{2016ApJ...818..113N}, or intensity mapping \citep[e.g., ][]{2020MNRAS.493.5854H}. To detect the \HI emission line beyond $z\sim 0.1$ is still a major challenge.

The new fully operational Five-hundred-meter Aperture Spherical radio Telescope \citep[FAST, ][]{2006ScChG..49..129N, 2011IJMPD..20..989N} with the tracking active reflector design \citep{1998MNRAS.301..827Q} can provide us the opportunity to probe \HI  in the higher redshift Universe with unprecedented sensitivity. For example, the upcoming Commensal Radio Astronomy FasT Survey (CRAFTS) project is expected to detect \HI  out to $z=0.35$ \citep{2019SCPMA..6259506Z}. As the largest filled-aperture radio telescope, FAST has been designed to achieve many challenging scientific goals, including hunting for pulsar, \HI  map of local galaxies, mapping the Milky Way central region, etc \citep[see the review of ][ and references therein]{Jiang2019, Jiang2020}. 

In this letter, we report the results of our pilot \HI  survey with FAST on a sample of four $z\sim0.05$ star star-forming galaxies taken from the Valpara\'iso ALMA/APEX Line Emission Survey \citep{Villanueva2017, Cheng2018}. They are among of the first extragalactic \HI detection cases observed by FAST in its commissioning phase. Throughout this paper, we assume a standard $\Lambda$CDM cosmology with $H_0=70\, \rm km\,s^{-1}\,Mpc^{-1}$, $\Omega_{\rm M} = 0.3$, and $\Omega_{\rm \Lambda} = 0.7$. All magnitudes are provided in the AB magnitude system \citep{1983ApJ...266..713O}.


\begin{figure}
    \centering
    \includegraphics[width=0.45\textwidth]{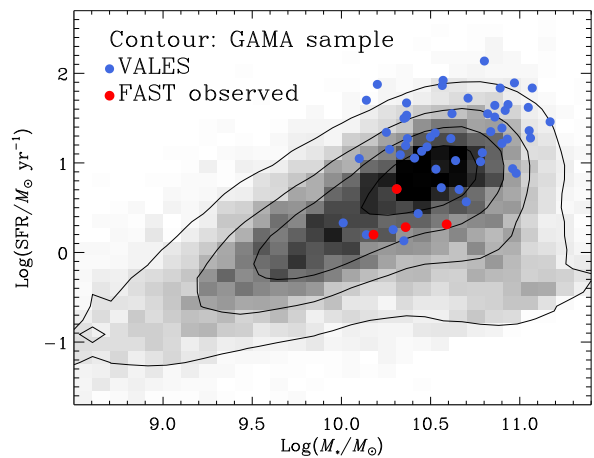}
    \caption{Main sequence of the low-$z$ galaxies (black contour and grayscale) and our ALMA detected VALES sample (blue dots). The FAST observed targets are shown as red dots.}
    \label{FAST_MS}
\end{figure}

\begin{figure*}
    \centering
    \includegraphics[width=0.45\textwidth]{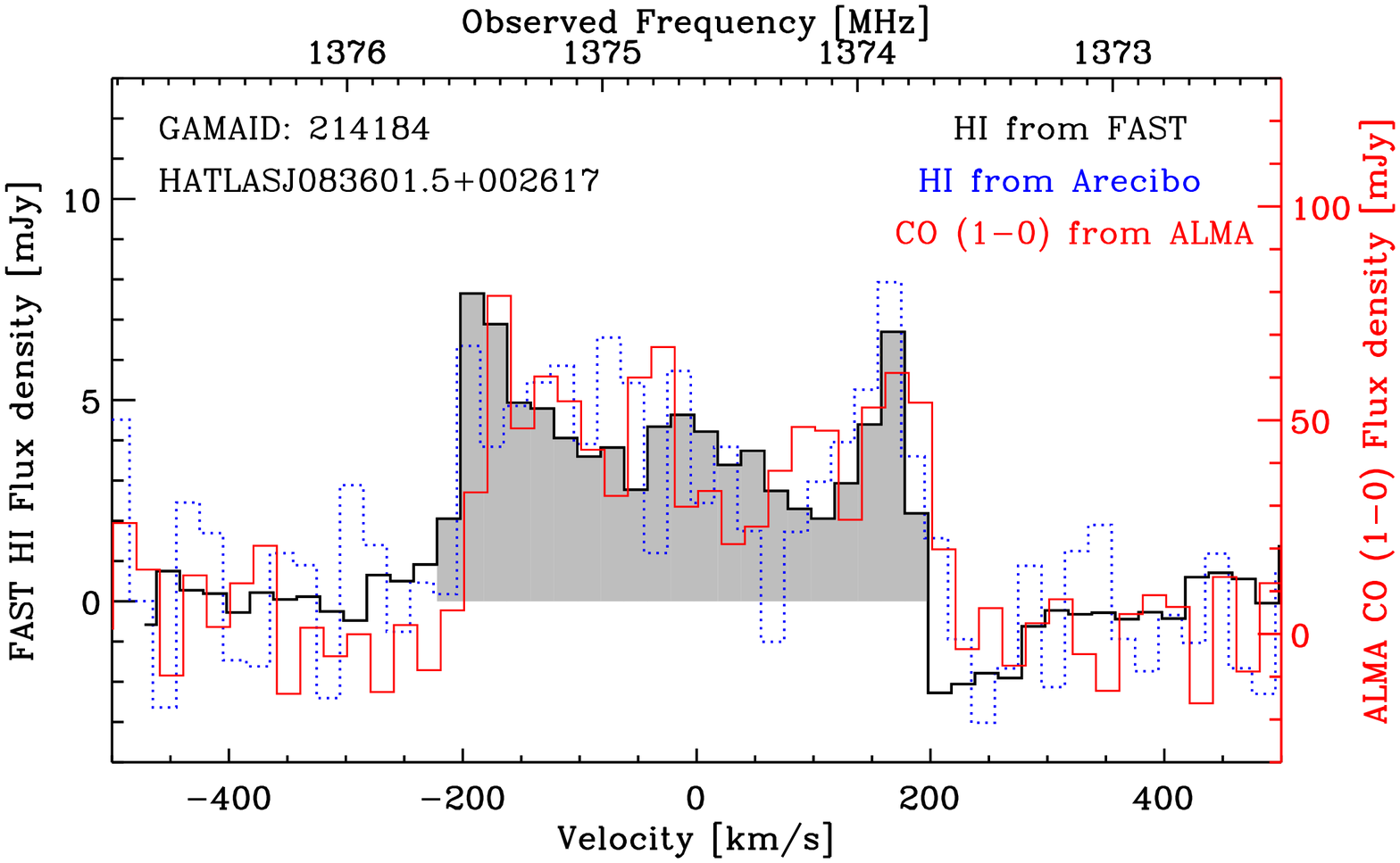}
    \includegraphics[width=0.45\textwidth]{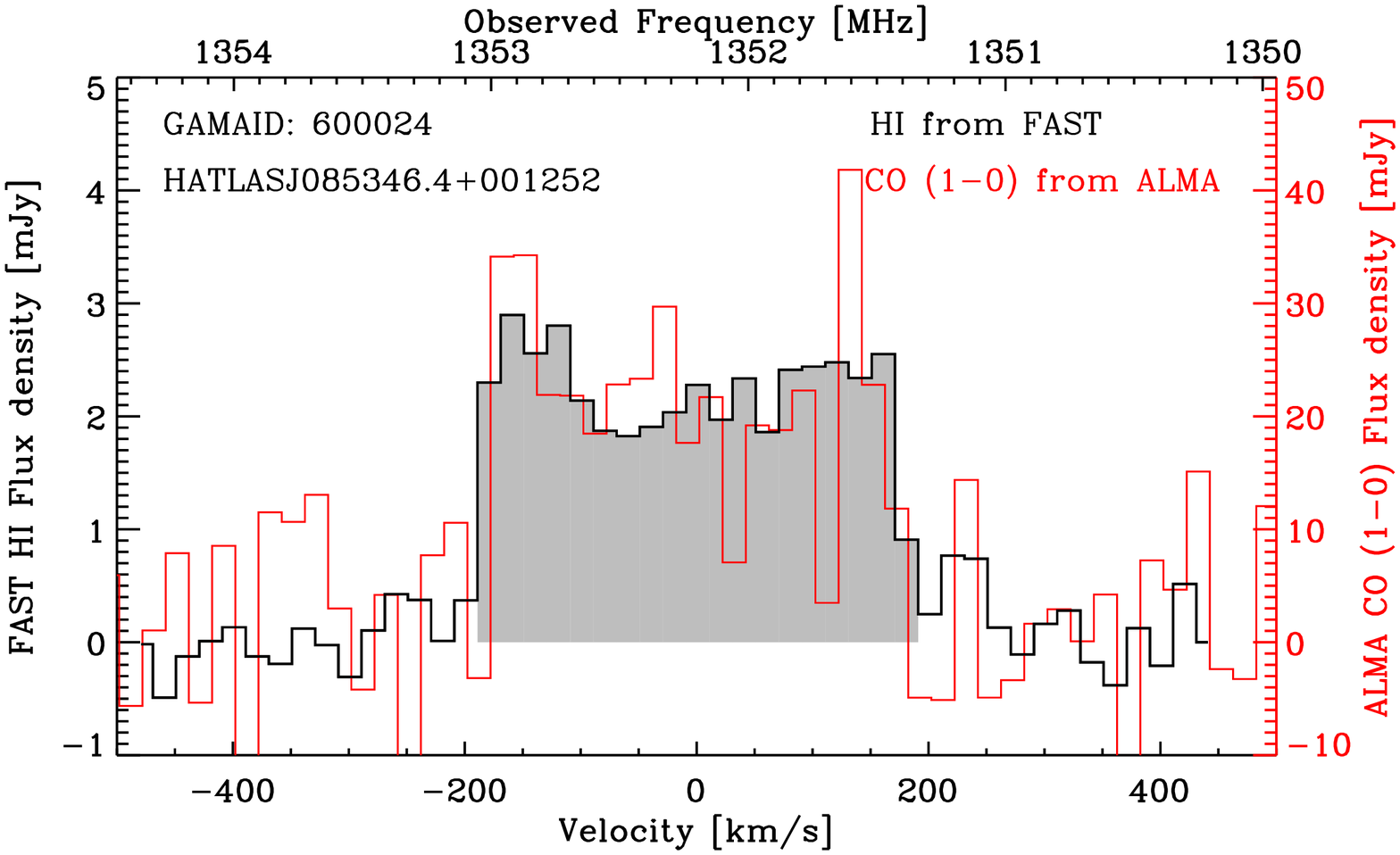}
    \includegraphics[width=0.45\textwidth]{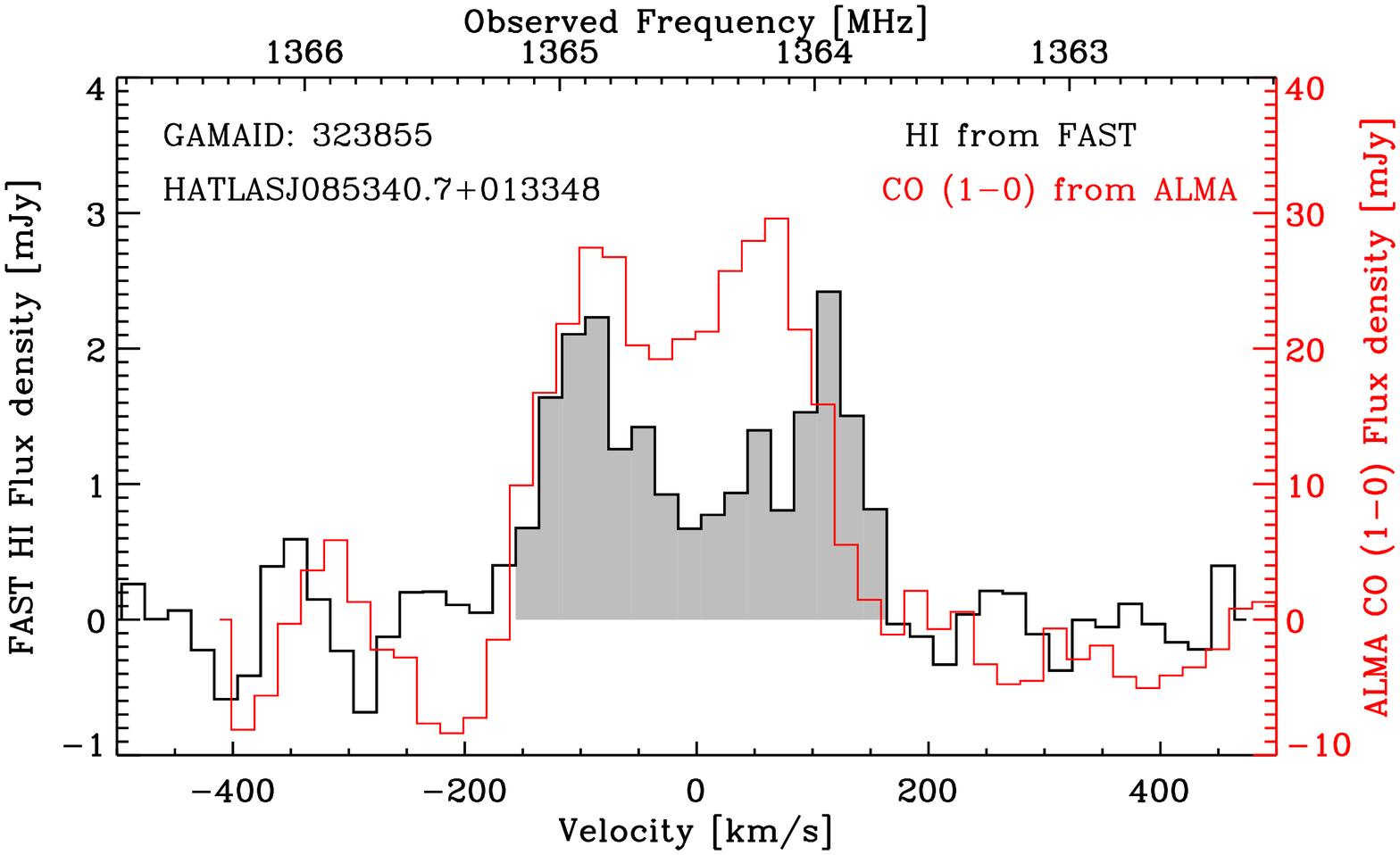}
    \includegraphics[width=0.45\textwidth]{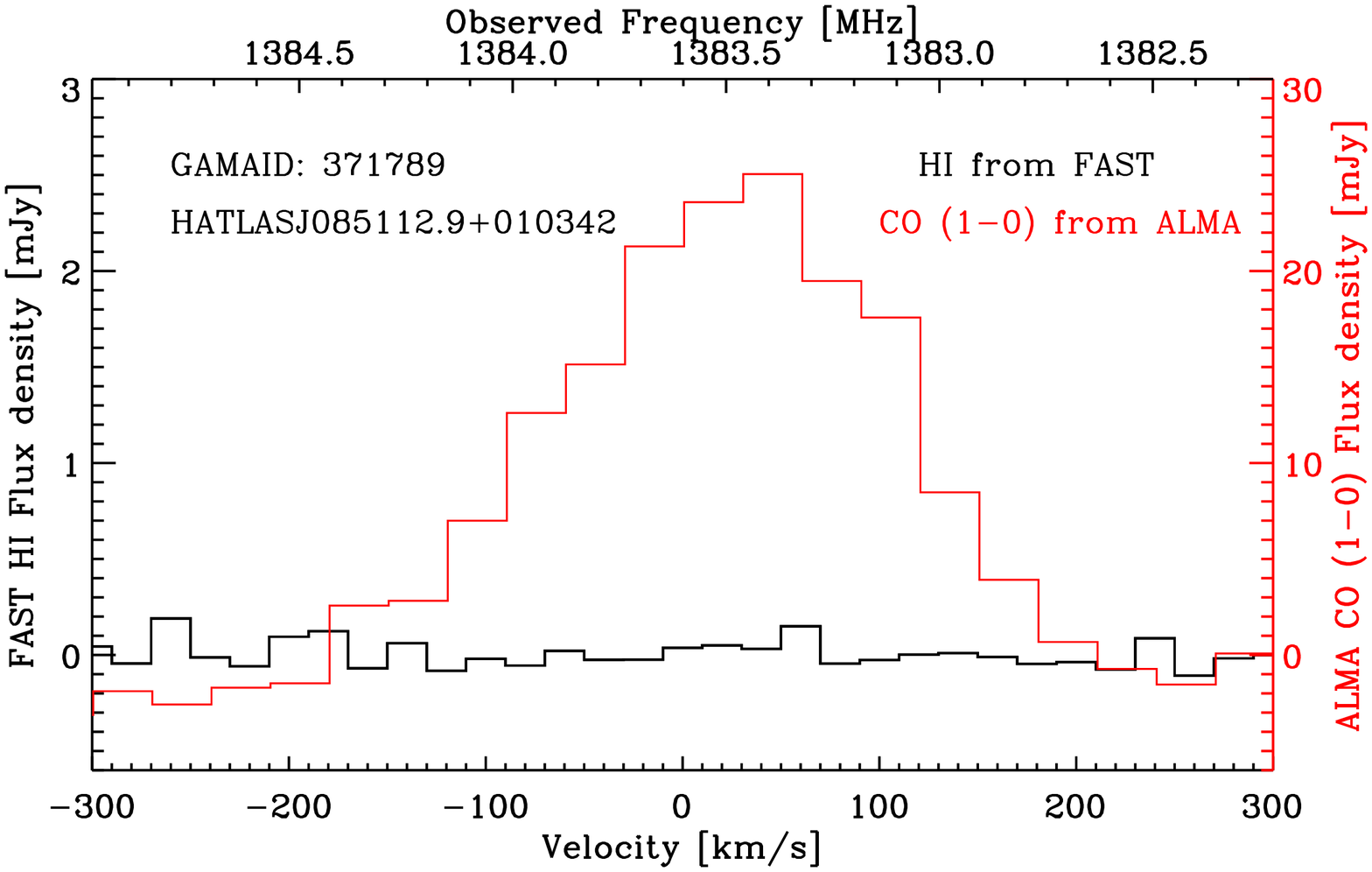}
    \caption{The black thick lines show the \HI  spectra obtained from FAST. The \HI  spectrum are rebined into 20 km/s resolution. The red dot lines are the CO (1-0) line from our previous ALMA survey. The scales of the CO emission lines are shown in the right y-axis. We adopt the optical spectroscopy redshift to derive the line velocity. We show the ALFALFA \HI  spectrum of the galaxy HATLASJ083601.5+002617 in blue color in the upper left panel. Effective on-target integration time of Arecibo telescope is about 48\,s. The edges of the emission lines are very sharp, and we obtain a consistency line width with the ALFALFA results. 
    The grey regions highlight the velocity channels we used to derive the \HI flux.
    }
    \label{HIspec}
\end{figure*}

\section{FAST observations}

\subsection{Sample selection}
The ongoing VALES project has targeted the low-$J$ CO transitions in 91 low-$z$ dusty star-forming galaxies taken from the {\it Herschel} Astrophysical Terahertz Large Area Survey \citep[$H$-ATLAS, ][]{2010PASP..122..499E}. The CO($J=1-0$) and CO($J=2-1$) emission lines have been observed using the Atacama Large Millimeter/sub-millimiter Array \citep[ALMA, ][]{Villanueva2017} and the Atacama Pathfinder EXperiment \citep[APEX, ][]{Cheng2018}, respectively. The VALES galaxies are selected from the equatorial GAMA fields, which provides extensive multi-wavelength coverage \citep{2009A&G....50e..12D}.

In this pilot work, we extract four galaxies previously detected in CO($J=1-0$) to target their \HI  emission line with FAST. To mitigate sensitivity effects, we focus on those at the lower redshift end of the VALES distribution. This is particularly useful to avoid the strong RFI seen at lower frequencies. The targeted sources have: (1) an expected \HI  flux higher than 0.5\,Jy\,km\,s$^{-1}$ as estimated using the \citet{Zhang2009} prescription, (2) they do not have bright  nearby galaxies  at similar redshift within one beam (2.9\,arcmin diameter), (3) their Declinations are in the range of 0 to 40\,deg, so that we can have a high collecting area (full 300m dish). Most of the VALES galaxies are massive star forming galaxies as shown in Fig. \ref{FAST_MS}. The observed four galaxies observed by FAST, shown as red dots, are local main sequence galaxies.

Aided by the previous coverage provided by ALFALFA \citep{Haynes2011,Haynes2018} in part of the GAMA fields, we include one source with ALFALFA measurements (HATLASJ083601.5+002617) to check consistency and the relative performance of the FAST telescope.

\subsection{Observation setup and calibration}

We were allocated 10 hours in L-band with the 19-Beam receiver during the FAST Commissioning Phase (Shared-risk Observing Proposal\footnote{\url{http://english.nao.cas.cn/ne2015/News2015/201902/t20190222_205600.html}} 2019A-012-S; PI: Cheng Cheng). 
The focal plane is covered by 19 beams, each one with a 2.9\,arcmin diameter and separated by 5.8\,arcmin from each other. 
Our observations were carried out in the tracking mode, utilizing only the central beam (M01).
Details for the FAST technical performance can be found in \citet{Jiang2020}. 

To mitigate the errors due to an unsteady baseline at the GHz frequency, we set the observations as 5\,min ON-target $+$ 5\,min OFF-target per iteration, and made from 3 to 8 iterations per target. The OFF-target pointing position was chosen at 5.8 arcmin distance, with no sources in the Focus of View (FoV) with a similar redshift (from SDSS spec-$z$ catalogue) as our targets. We use the flux from the OFF pointing to estimate the background, and thus the ON-OFF spectra is the flux from our target.

We calibrate the flux by the noise diode with known antenna temperature that injected into the receiver during the observation \citep{Jiang2020}. Detail data reduction and calibration process are briefly described in Sec. \ref{datapipeline}.

\begin{table*}
\caption{FAST observation results. $W_{300}= \rm Width / 300$ km/s is used to estimate the \HI flux of HATLASJ085112.9+010342.}
\tiny
\label{table:1}
\centering          
\begin{tabular}{ccccccc}     
\hline\hline       
HATLAS ID  & GAMAID  & $t_{\rm int}$ & rms @ 1.7\,km\,s$^{-1}$    & \HI  flux           & \HI  FWHM         & Peak to peak \\
           &         &    min        & mJy\,beam$^{-1}$           & mJy\,km\,s$^{-1}$ & km\,s$^{-1}$ & km\,s$^{-1}$ \\
\hline
HATLASJ083601.5+002617 & 214184  & 5      & 2.61  & 1696.5 $\pm$ 188.1 & 388$\pm$10 & 361.1 \\
HATLASJ085346.4+001252 & 600024  & 10     & 1.59   & 812.6 $\pm$ 85.2  & 364$\pm$10 & 328.1  \\
HATLASJ085340.7+013348 & 323855  & 15     & 0.97   & 410.4 $\pm$ 42.3 &  341$\pm$10 & 229.6  \\
HATLASJ085112.9+010342 & 371789  & 20     & 0.73 &  $< 1095\, W_{300}$ &  &      \\
\hline    
\end{tabular}

\end{table*}
\begin{table*}
\caption{Target properties.  The lower limit of the $\log M_{\rm bary}$ is estimated by $M_{\rm H_2} + M_*$}
\tiny
\label{table:2}
\centering          
\begin{tabular}{ccccccccc}     
\hline\hline       
GAMAID &  $z_{\rm spec}$   & $\log\,M_{\rm H\textsc{i}}$ & log\,$L'_{\rm CO}$  &  CO FWHM    & $\log\,M_*$     & $ \log M_{\rm dyn}^{\rm CO} $ & $ \log M_{\rm bary}$ & $ \log M_{\rm dyn}^{\rm H\textsc{i}}$ \\
       &                   &          $M_\odot$          & K km/s pc$^2$       &  km/s       &  $M_\odot$      &  $M_\odot $                   & $ M_\odot     $       & $ M_\odot     $  \\
\hline
214184 &  0.0332           & 9.91 $\pm$  0.12            & 9.02 $\pm$ 0.02     &  391.2      & 10.59 $\pm$ 0.1 & 10.75$\pm$0.02                & 10.68$\pm$0.02 & 11.96 $\pm$ 0.05   \\
600024 &  0.0504           & 9.96 $\pm$  0.11            & 8.88 $\pm$ 0.01     &  349.9      & 10.31 $\pm$ 0.1 & 10.89$\pm$0.05                & 10.51$\pm$0.04 & 11.90 $\pm$ 0.03   \\
323855 &  0.0410           & 9.48 $\pm$  0.11            & 8.79 $\pm$ 0.02     &  342.4      & 10.36 $\pm$ 0.1 & 10.77$\pm$0.08                & 10.56$\pm$0.03 & 11.30 $\pm$ 0.03  \\
371789 &  0.0266           & $< 9.53 + \log W_{300}$ & 8.30 $\pm$ 0.07 &  197.2 & 10.14 $\pm$ 0.1 & 9.89$\pm$0.24                 & $>$10.17       &     \\
\hline
\end{tabular}
\end{table*}

\section{Results}

The noise levels reached for each target are listed in Table~\ref{table:1}, while the \HI  spectra are shown in Fig.~\ref{HIspec}.  For a 5\,min on-target integration, we got the noise level about 2.6\,mJy\,beam$^{-1}$ at 1.7\,km\,s$^{-1}$ velocity resolution bin. The rms declines with the on-target integration time, but not in the manner of $\propto 1/\sqrt{t}$, which might be caused by the {pointing jitter in the commissioning stage, or weak interference we have not been aware of}. We detect the \HI emission line in three galaxies. For the \HI undetected galaxy, we estimate the flux upper limit as $5\times$rms$\times 300 W_{300}$ mJy km/s, where $W_{300} = \rm Width/300$ km/s.
All three detected spectra show the double-horn pattern with a typical Full Width Half Maximum (FWHM) of about 300\,km\,s$^{-1}$. We re-sample the \HI  spectrum into $\Delta V = 20$\,km\,s$^{-1}$ channels, and derive the integrated \HI flux by summing over the velocity channels within Full Width at Zero Intensity (FWZI), which includes the channels at both line wings with the channel flux higher than the rms. We highlight the velocity channels that used to derive the \HI flux in grey color in Fig. \ref{HIspec}).

We derive the \HI  mass following \citet{Giovanelli2015}:
\begin{equation}
\frac{M_{\rm H\textsc{i}}}{M_\odot} = \frac{2.35\times10^5 D_{\rm Mpc}^{2}}{1+z} \int_{\rm FWZI} S(V) dV,
\end{equation}
where the $S(V)$ is the flux density in unit of Jy\,beam$^{-1}$. The uncertainty is estimated by the $\sigma^2 = (\sqrt{N_{\rm channel}} \Delta V\, \times {\rm rms})^2 + (10\% \times \rm flux)^2$, where $N_{\rm channel}$ is the number of integration channels, the $10\% \times \rm flux$ would account for the uncertainty of the flux calibration. The \HI mass or mass upper limit of our targets are listed in Table~\ref{table:2}.

\begin{figure}
    \centering
    \includegraphics[width = 0.47\textwidth]{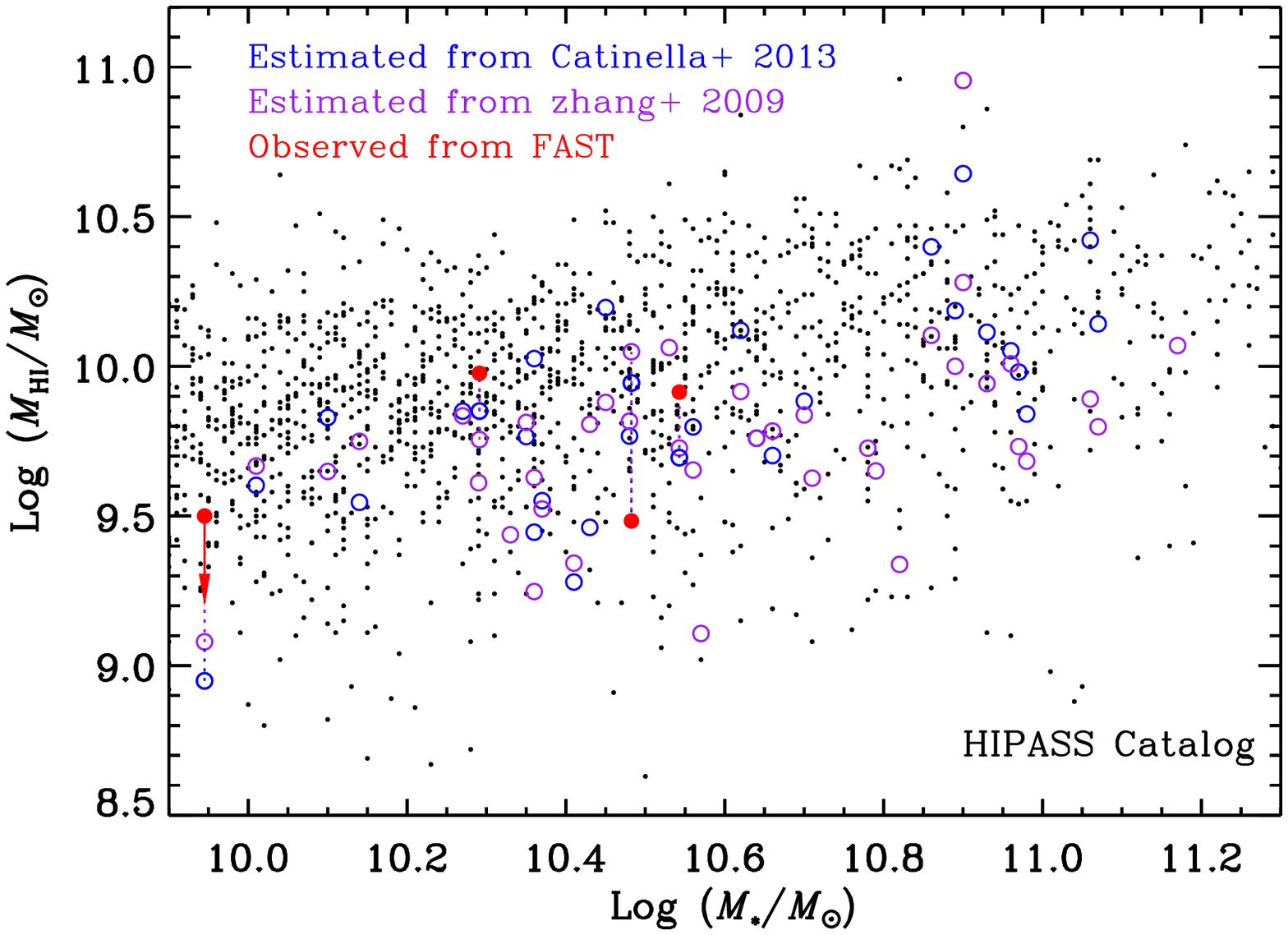}
    \caption{ Predicted \HI  masses of our VALES sample based on the empirical relations
    in \citet[][ in purple circles]{Zhang2009} and in \citet[][ in blue circles]{2013MNRAS.436...34C}. The black dots are the galaxies from HIPASS \citep{2018ApJ...864...40P}. The red dots show the results of the FAST observation. We link the predicted and the observed \HI  mass by dot lines. We show the FAST non-detected target by upper limits.}
    \label{VALESHImass}
\end{figure}

\section{Discussion}

\subsection{Comparison with a previous ALFALFA result}

For HATLASJ083601.5+002617, in upper left panel of Fig. \ref{HIspec} we show the results from both FAST and Arecibo telescopes. The ALFALFA survey is in the drift scan mode, and the effective on the target integration time for the ALFALFA spectrum is about 48\,s \citep{Haynes2011}. The observed spectra show consistent line profiles. The integrated flux of the Arecibo spectrum given in \citet{Haynes2018} is $1.68 \pm 0.14$ mJy\,km\,s$^{-1}$, and the FWHM is 390 $\pm$ 19 \,km\,s$^{-1}$, both values consistent with our results (see Table~\ref{table:1}). This indicates a reasonable flux calibration for the FAST 19-beam receiver in this commissioning phase experiment. In terms of noise levels, at 20\,km\,s$^{-1}$, we reach 0.8\,mJy\,beam$^{-1}$ for 5 min integration time, corresponding to about 2 mJy\,beam$^{-1}$ for 48s integration, while the rms given in ALFALFA \HI catalogue is 2.89 mJy\,beam$^{-1}$ at 10 km/s resolution, which is about 2 mJy\,beam$^{-1}$ at 20 km/s resolution. Thus it appears that, in the commissioning phase, the FAST already reached a sensitivity similar to that of ALFALFA.

\subsection{The {\rm \HI}  gas mass}

Previous studies have shown empirical correlations to derive the \HI  gas mass from the global optical properties such as the color, surface brightness, and the stellar mass densities \citep[e.g., ][]{Zhang2009, 2013MNRAS.436...34C, 2018ApJ...864...40P} with typical scatters about 0.5\,dex.
In Fig.~\ref{VALESHImass}, we show the observed and estimated \HI  mass of our VALES sample based on the relations given by
\citet{2013MNRAS.436...34C}:
\begin{equation}
    \log(M_{\rm H\textsc{i}}/M_*) = -0.338 \log \Sigma_* - 0.235 (NUV-r) + 2.908,
\end{equation}
where $\Sigma_* = M_*/\pi R_{\rm opt}^2 \, [M_\odot/\rm kpc^2]$ is the stellar mass surface density, $ R_{\rm opt}$ is the $r$ band half light radius, and the relation from \citet{Zhang2009}:
\begin{equation}
    \log(M_{\rm H\textsc{i}}/M_*) = -1.73238 (g-r) + 0.215182 \mu_i - 4.08451,
\end{equation}
where $\mu_i$ [mag/arcsec$^2$] is the SDSS i band average surface brightness. 
We adopt the stellar mass, radius and the photometry from GAMA galaxy structure catalogue \citep{Kelvin2012}
and $H$-ATLAS catalogue \citep{2016MNRAS.462.3146V}. We also show the observed results from \HI  Parkes All-Sky Survey catalogue \cite[HIPASS, ][]{2018ApJ...864...40P} in Fig. \ref{VALESHImass} as a comparison.

In \citet{Villanueva2017} we show that the VALES sample stands for the galaxy population from star-forming (for the $z\sim 0.05$ galaxies) to the starburst \citep[mainly at $z>0.1$, see the Fig. 1 of ][]{Villanueva2017}. We found the VALES sample follow the Kennicutt-Schmidt relation based on the H$_2$ gas mass from  CO($J=1-0$) observations, and the \HI  gas mass derived from the empirical relation \citep{Zhang2009}. Our pilot survey results show a consistency between estimated and observed \HI mass, suggesting that the star-forming galaxies at redshift 0.05 in VALES sample may still follow the Kennicutt-Schmidt relation. More \HI observations of the VALES galaxies by FAST are still ongoing.

\subsection{Dynamical masses}
In \citet{Molina2019}, we presented a dynamical analysis to 39 VALES galaxies using the spatial extension of the CO($J=1-0$) emission. For the three \HI  detected galaxies in this work, the dynamical study shows that the CO rotation curves are still increasing or just about to turn to flat rotation curve with the radius \citep[see the Appendix A of ][]{Molina2019}. While comparing the \HI  and CO($J=1-0$) emission line spectra, we find that the CO($J=1-0$) line profiles in Fig. \ref{HIspec} show a sharp decline at larger velocities, and a similar line width and profile as the \HI  spectra (see the CO FWHM in 
Table~\ref{table:2}). The agreement of the CO and \HI  line width suggests that the observed CO($J=1-0$) emission might extend up to the region where the rotation curve is flat \citep{1992ApJ...393..530D, 1992PASJ...44L.231S, 1994A&A...283...21S, 2016AJ....152...51D, Tiley2016}.

We can roughly estimate the dynamical mass based on the \HI  FWHM, and \HI  radius ($R_{\rm H\textsc{i}}$) by $M_{\rm dyn}^{\rm H\textsc{i}} = ({\rm FWHM/2/\sin \theta})^2 R_{\rm H\textsc{i}} / G$, where $\theta$ is the \HI  inclination angle, $R_{\rm H\textsc{i}}$ is the \HI  radius. Since for galaxies the \HI size is tightly correlated to the \HI mass \citep{2016MNRAS.460.2143W, Stevens2019}, the $R_{\rm H\textsc{i}}$ can be estimated from $M _{\rm H\textsc{i}}$ using the relation derived by \citet{2016MNRAS.460.2143W}. For the \HI  mass range of $ 9.5 < \log(M_{\rm H\textsc{i}}/M_\odot) < 10 $, the \HI  radius is in the range of 16.5 to 30\,kpc \citep[See Eqn. 2 or Fig. 1 in][]{2016MNRAS.460.2143W}, which is much larger than the CO radius (see the red contours in Fig. \ref{img}). We do not have inclination information of the \HI  gas. But if we assume the Ks band image and \HI  have a similar inclination angle \citep{Molina2019}, we can roughly estimate the $M_{\rm dyn}^{\rm H\textsc{i}}$(see Table~\ref{table:2}). The $M_{\rm dyn}^{\rm H\textsc{i}}$ is about one order of magnitude larger than the baryon mass ($M_{\rm bary} = M_{\rm H\textsc{i}} + M_{\rm H_2} + M_*$, also listed in Table~\ref{table:2}. For HATLASJ085112.9+010342, which has no \HI detection, we estimate the lower limit of the baryon mass as $M_{\rm bary}^{\rm lower\, limit} = M_{\rm H_2}+M_*$). The \HI  radii of our targets are much larger than the optical and CO radii, yielding a larger $M_{\rm dyn}^{\rm H\textsc{i}}$. The larger $M_{\rm dyn}^{\rm H\textsc{i}}$ may suggest that \HI  mainly traces the dynamical mass within a larger radius where the dark matter start to dominate the gravitational potential.

On the other hand, our previous work also derived the dynamical mass within $2\times r_{1/2 \rm CO}$ from CO velocity map \citep{Molina2019}, where the half light CO radius ($r_{1/2 \rm CO}$) is about 4 kpc. We list the $M_{\rm dyn}^{\rm CO}$ in Table~\ref{table:2} and we can see a good consistency between the $M_{\rm dyn}^{\rm CO}$ and the baryon mass. Considering the higher dynamical masses derived from the \HI  data, this implies that CO dynamics is restricted, as expected, to the gravitational potential of the central regions in galaxies. Therefore, it might be possible to trace dark matter halos using both CO and \HI  observations. We show the optical image, CO contours as well as the \HI spectra of our targets in Sec. \ref{colorimg}. The target HATLASJ085340.7+013348 has a clumpy CO morphology, which might be the reason of the relatively higher CO to \HI flux ratio in Fig. \ref{HIspec}.

\section{Conclusion} \label{sec:conclusion}

We report on some of the first extragalactic \HI  line observations made during the commissioning phase of the FAST 19-beam receiver, of four star-forming galaxies at $z\simeq 0.05$ taken from the VALES survey. These are among the first extragalactic \HI  detection results during the FAST commissioning phase stage. Using 5min/5min ON/OFF pointing observations, we reached an rms of 2.6\,mJy\,beam$^{-1}$ at a spectral resolution 1.7\,km\,s$^{-1}$. We detected three out of the four observed galaxies. One of our targets was detected previously by the Arecibo ALFALFA survey, with results consistent with ours. The observed \HI  masses are consistent with values estimated using previously determined empirical relations \citep{Zhang2009, 2013MNRAS.436...34C}, with a scatter of about 0.5 dex. We find the width of the \HI  emission is similar to the CO($J=1-0$) width revealed by ALMA, suggesting that ALMA observations already observed the flat rotation curve from the galaxy outskirts. The dynamical mass that estimated from \HI  ($M_{\rm dyn}^{\rm H\textsc{i}}$) is an order of magnitude higher than the baryon mass ($M_*+M_{\rm H_2}+M_{\rm H\textsc{i}}$) and the dynamical masses derived from CO observations ($M_{\rm dyn}^{\rm CO}$), implying that the dynamical mass traced by \HI  would be more dominated by dark matter halo. 

\begin{acknowledgements}
We thank the referee for carefully reading and for patiently providing constructive comments that helped us to improve the quality of this paper.
C.C. appreciates the kindness help from FAST, especially to Lei Qian, Ningyu Tang, Jing Tang and Zheng Zheng, and helpful discussions about FAST observation configuration and data reduction with Pei Zuo, Niankun Yu and Lizhi Xie. This work made use of the data from FAST (Five-hundred-meter Aperture Spherical radio Telescope). FAST is a Chinese national mega-science facility, operated by National Astronomical Observatories, Chinese Academy of Sciences. C.C. is supported by the National Natural Science Foundation of China (NSFC), No.\ 11803044, 11673028. J.H. is supported by NSFC, No.\ 11933003. This work is sponsored (in part) by the Chinese Academy of Sciences (CAS), through a grant to the CAS South America Center for Astronomy (CASSACA). E.I.\ acknowledges partial support from FONDECYT through grant N$^\circ$\,1171710. D.W. is supported by NSFC, No.\ U1931109, 11733006. T.M.H. acknowledges the support from the Chinese Academy of Sciences (CAS) and the National Commission for Scientific and Technological Research of Chile (CONICYT) through a CAS-CONICYT Joint Postdoctoral Fellowship administered by the CAS South America Center for Astronomy (CASSACA) in Santiago, Chile. This work was supported by the National Science Foundation of China (11721303, 11991052) and the National Key R\&D Program of China (2016YFA0400702). CKX acknowledges support from the National Key R\&D Program of China No. 2017YFA0402704 and National Natural Science Foundation of China No. 11873055 and No. 11733006. This paper makes use of the following ALMA data: ADS/JAO.ALMA\#2013.1.00530.S. ALMA is a partnership of ESO (representing its member states), NSF (USA) and NINS (Japan), together with NRC (Canada), MOST and ASIAA (Taiwan), and KASI (Republic of Korea), in cooperation with the Republic of Chile. The Joint ALMA Observatory is operated by ESO, AUI/NRAO and NAOJ. In addition, publications from NA authors must include the standard NRAO acknowledgement: The National Radio Astronomy Observatory is a facility of the National Science Foundation operated under cooperative agreement by Associated Universities, Inc.

\end{acknowledgements}

%
    \bibliographystyle{aa} 
    \bibliography{aanda.bib}{} 
%

\begin{appendix}
\onecolumn
\section{Data reduction and flux calibration}\label{datapipeline}

\begin{figure*}
    \centering
    \includegraphics[width=0.99\textwidth]{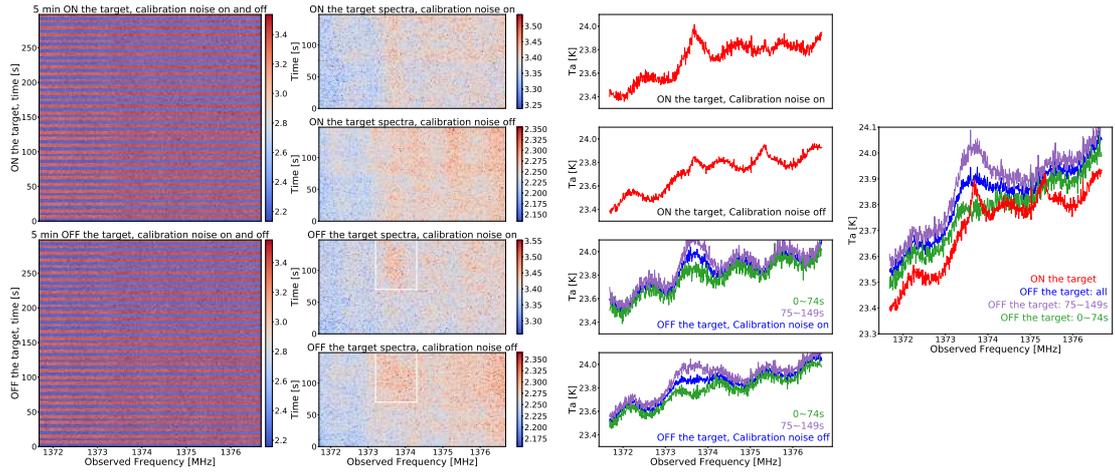}
    \caption{Data reduction pipeline: {\bf Left two panels:} Each panel shows the raw spectra for the 5-min ON- or OFF-target positions (upper and lower panel respectively). The x-axis is the observed frequency, 
    the y-axis shows the sampling time of the spectra, with a sampling rate of once every 1.006632926s (0.993411 Hz). The signal intensity in individual spectra (300 of them in each panel) is represented by colors, with the scale illustrated by the color bar on the right side of each panel. Note that those spectra in ``red stripes" were sampled when calibration noise was on, and those in ``blue stripes" were sampled when the calibration noise was off (see text).
    Unit of the color bar is digital number / $10^{12}$ from ADU, which needs to be calibrated into temperature unit.
    {\bf middle left panels:} We split the raw spectra into four cases: ON/OFF the target with calibration noise on/off. 
    {\bf Middle right panel:} The average spectra of the four classes. We show the total background in blue color, and the background from the first and rest 74 seconds (green and purple color) in the lower two panels. The green line show a lower temperature than the purple line, implying a increasing system temperature. 
    {\bf Right panel:} Spectra of the ON and OFF the target.}
    \label{pipeline}
\end{figure*}

Fig.~\ref{pipeline} shows the data reduction flowchart. The data is reduced in chunks of 10\,min. Each chunk includes 300 spectra for the ON target and 300 spectra red for the OFF target (background). If the redshifted \HI  line suffers from temporal RFI, we simply ignore these 10\,min data. The left two panels in Fig.~\ref{pipeline} are the spectra we obtained in the 5\,min ON (upper panel) and 5\,min OFF the target (bottom panel). 

A noise diode is used for FAST signal calibration \citep{Jiang2020}. When it is `on', it injects a noise of known temperature to the receiver. The noise diode is switched between `on' and `off' repeatedly during the whole FAST observation. Noise `on' and `off' are in the periods twice the signal sampling period. We extract one spectrum per 1.006632926s. Half of the spectra contains noise with known temperature from noise diode. For a given target, comparing the spectra with noise diode `on' and `off', we can derive the intensity (i.e. antenna temperature) of these spectra \citep{Jiang2020}.

Our spectra can be classified in four classes, ON target with noise diode `on' and `off' ($f_{\rm ON}^{\rm cal\, on}$ and $f_{\rm ON}^{\rm cal\, off}$), and OFF target with noise diode `on' or `off' ($f_{\rm OFF}^{\rm cal\, on}$ and $f_{\rm OFF}^{\rm cal\, off}$). These four classes are shown in the middle-left panels of Fig.~\ref{pipeline}. Note these have not been yet calibrated in flux density. 

The flux calibration is done via the following two formulas: $T_{\rm ON}^{\rm cal\, off}/f_{\rm ON}^{\rm cal\, off} =  T_{\rm noise}/(f_{\rm ON}^{\rm cal\, on}-f_{\rm ON}^{\rm cal\, off})$, and $(T_{\rm ON}^{\rm cal\, on}+T_{\rm noise})/f_{\rm ON}^{\rm cal\, on} = T_{\rm noise}/(f_{\rm ON}^{\rm cal\, on}-f_{\rm ON}^{\rm cal\, off})$, where $T_{\rm noise}$ is the known noise temperature in Kelvin, the $T_{\rm ON}^{\rm cal\, on}$ and $T_{\rm ON}^{\rm cal\, off}$ are the calibrated antenna temperatures of the signal data, with noise diode `on' and `off' \citep[see the Sec. 3.1 of ][]{Jiang2020}. The average calibrated spectra of four classes are shown in the middle right panels of Fig.~\ref{pipeline}. These are inverse-variance weighted means of individual spectra, and the rms error is estimated as $\sigma \propto T_{\rm sys}/\sqrt{\Delta V t}$, where the system temperature $ T_{\rm sys} \sim 20\; $K, the channel width $\Delta V=1.7\;$ km\; s$^{-1}$, and the $t$ here is the integration time. The right panel of Fig.~\ref{pipeline} shows the weighted average spectra temperature (Ta) of the ON and OFF targets in the units of Kelvin.

We find a clear increment of the noise as we move to higher frequencies, starting at $\sim$ 1373.5\,MHz to 1374.5\, MHz (see white boxes in Fig.~\ref{pipeline}), indicating an unstable background noise. In order to assess this quantitatively, we derive additional mean spectra for the OFF position in two time intervals: the first 0$\sim$74\,s and the second 75$\sim$147\,s separately. We compare the mean OFF spectra of different time intervals and find that the background temperature increases by $\sim$ 0.1\,K during the 5 min integration, with a larger increase about 0.3\,K at 1373.5 MHz (purple and green color curves in the middle right panel of Fig. \ref{pipeline}). In the right panel of Fig.~\ref{pipeline}, we can also see that the target OFF spectrum (blue line) is about 0.15\,K higher than the target ON spectrum (red). 

We adopt a 14.86\,K\,Jy$^{-1}$ conversion  factor, which has an accuracy of the order of 10\% \citep{Jiang2020}.

\begin{figure*}
    \centering
    \includegraphics[width=0.99\textwidth]{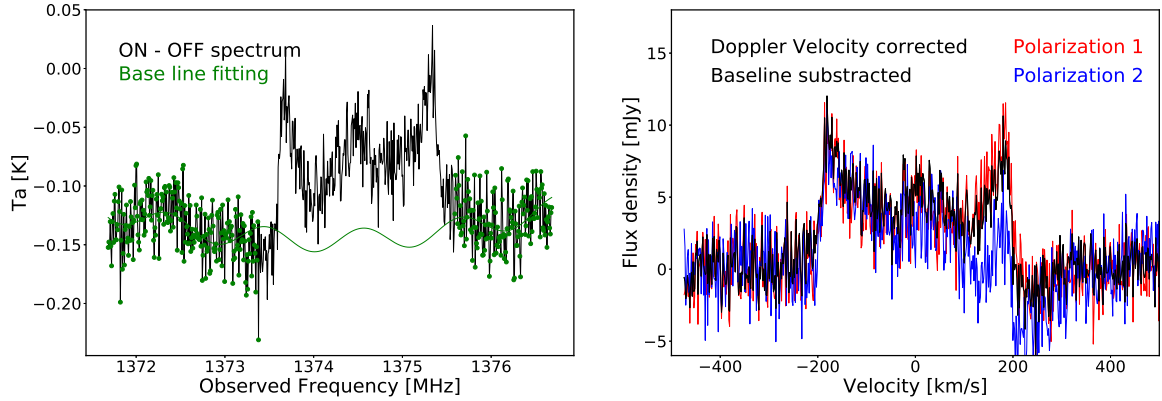}
    \caption{Flux calibration: {\bf left panel:} We fit the baseline from the data shown in green dots with the function $A*\sin (k\,x+b) + m \, x + n$ to account for the stand wave. {\bf right panel:} We correct the Doppler velocity and convert the antenna temperatures to flux intensities in mJy. \HI  emission from galaxies should have no polarization. As a consistency check, we also reduce the spectrum including only one polarization, and show the results in blue and red colors.}
    \label{calibration}
\end{figure*}

We subtract the baseline using model-fit based on a sinusoidal plus a linear function, where the sine function would account for the standing wave, and the linear function would represent the trend of the baseline between 1372 MHz and 1376 MHz. As an example, the final steps of the data reduction of a target are illustrated in Fig. \ref{calibration}.
The generation of the final spectrum can be seen in the left panel of Fig. \ref{calibration}. The minus temperature is caused by the increasing background noise. We correct the Doppler velocity and convert the spectrum velocity into kinematical Local Standard of Rest, and show the final spectrum in the right panel of Fig. \ref{calibration}.

Each FAST spectrum includes the spectra with two polarizations individually. Since \HI  emission from galaxies should not be polarized, we inspected both polarizations separately to check if the rms or detections are consistent. The right panel of Fig. \ref{calibration} shows the two polarizations in blue and red colours. We do not find significant differences in polarisation, so we simply combined the spectra of both polarizations together during the data reduction. 


\section{Fake color image of our targets}\label{colorimg}

\begin{figure*}
    \centering
    \includegraphics[width = 0.46\textwidth]{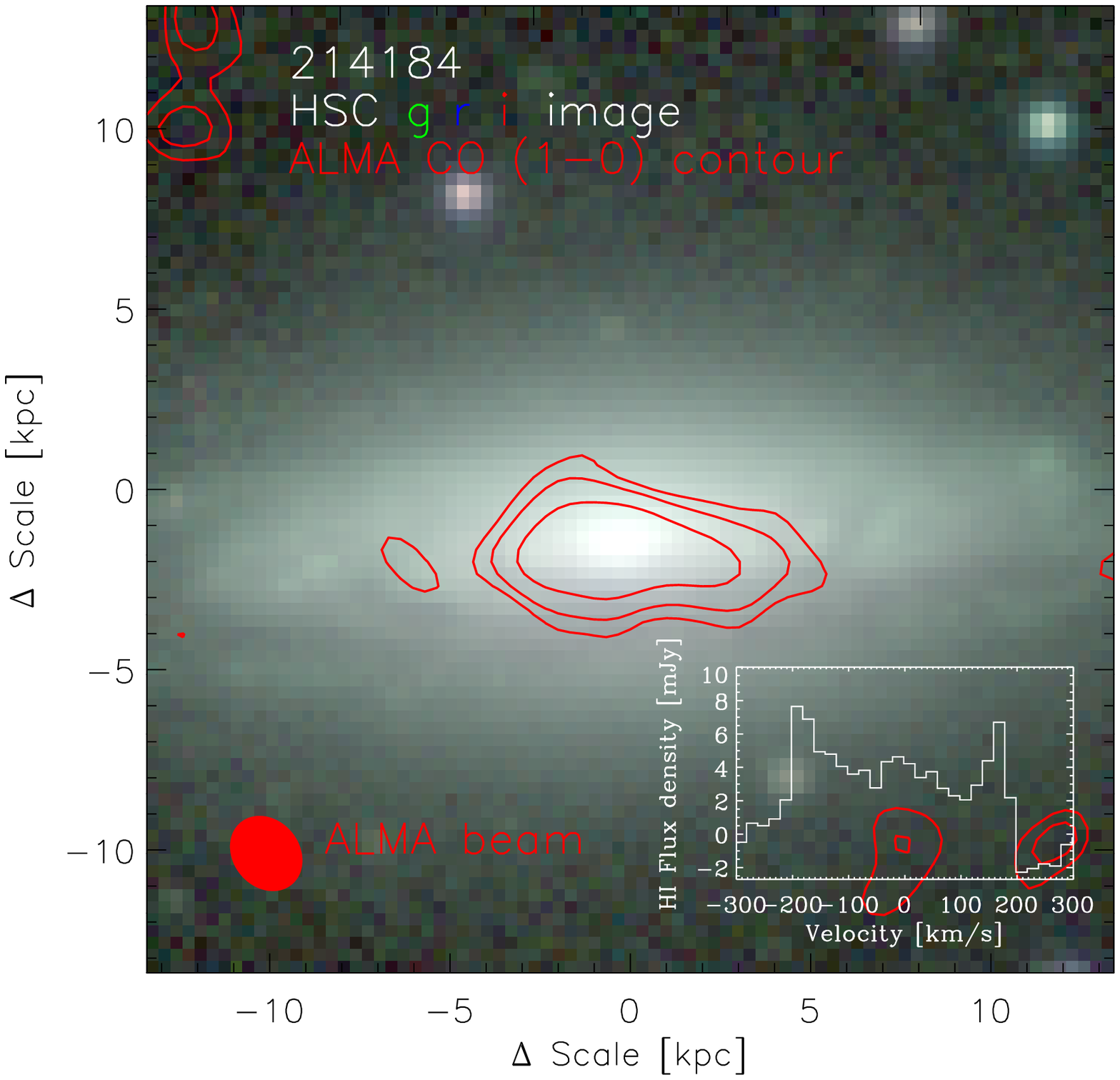}
    \includegraphics[width = 0.46\textwidth]{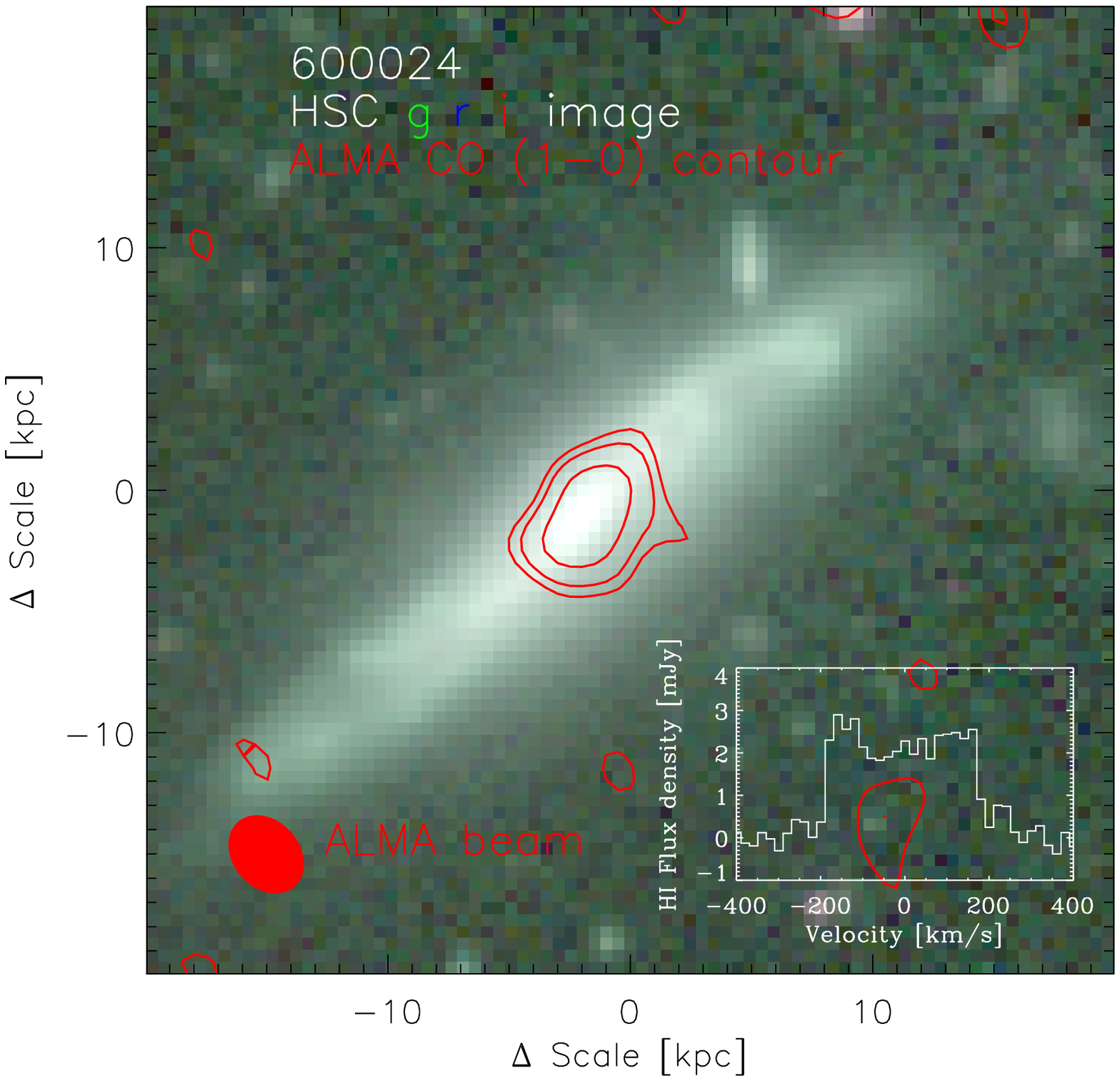}
    \includegraphics[width = 0.46\textwidth]{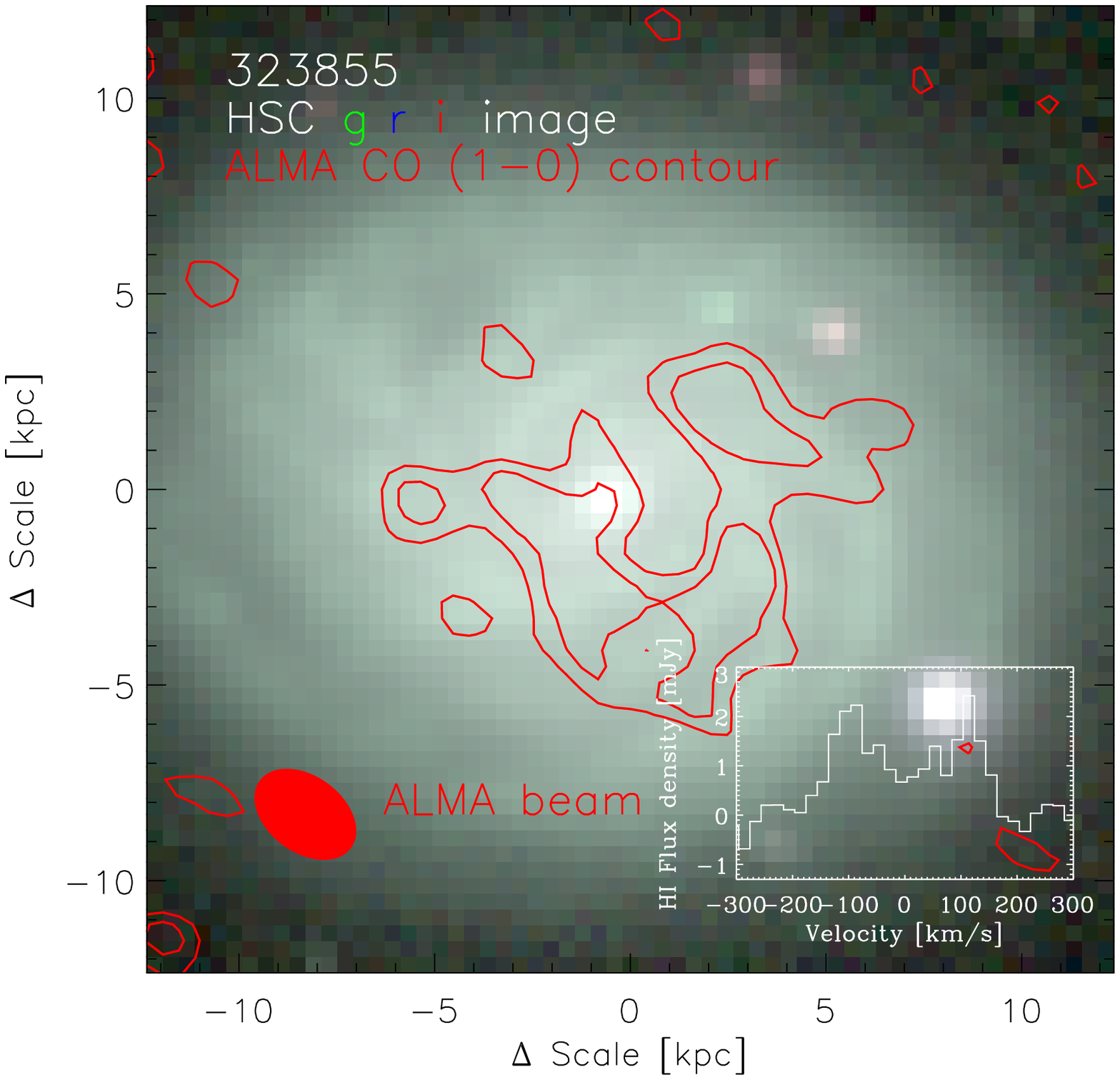}
    \includegraphics[width = 0.46\textwidth]{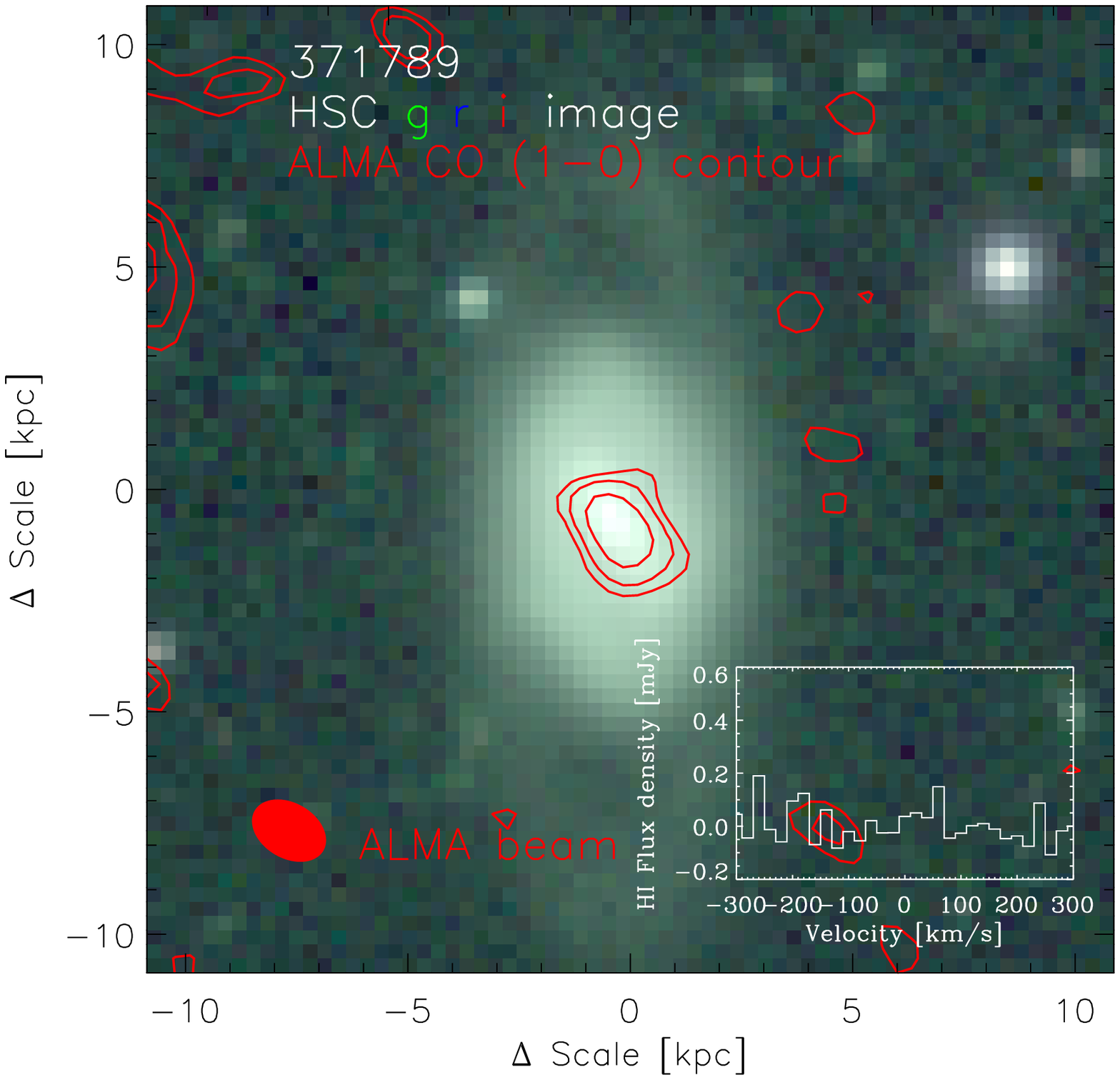}
\caption{Color images composed g, r, i band images from the HSC survey wide fields \citep{2019PASJ...71..114A}, as well as the red contour of ALMA CO (1-0) moment 0 map. The contours are shown in [2, 3, 5]$\times$ rms. We also show the \HI  spectrum in 20 km/s velocity bin at the lower right corner of each panel. The FAST beam size (2.9 arcmin) is corresponding to about 130 kpc at redshfit 0.05, much larger than the scale of the images.
}\label{img}
\end{figure*}

We show the CO contour and the optical images of our targets in Fig. \ref{img}. we can see that the CO morphology is compact in the galaxy center for HATLASJ085346.4+001252, HATLASJ083601.5+002617, and HATLASJ085112.9+010342. And for the galaxy HATLASJ085340.7+013348, the CO morphology is clumpy. We also see that the H$_2$ to \HI  flux ratio for HATLASJ085340.7+013348 is about 15, while the rest two targets have CO to \HI  flux ratios about 10. This might suggest a more efficient formation of H$_2$ molecules in a nuclear region, and a possible connection between CO morphology and the CO to \HI  flux ratio.

\end{appendix}

\end{document}